\documentclass[12pt]{iopart}

\usepackage[numbers,sort&compress]{natbib}

\usepackage[dvipsnames]{xcolor}
\usepackage{graphicx} 
\usepackage{color} 
\usepackage{epsfig}
\usepackage{hyperref} 
\usepackage[english]{babel}

\begin{document}
\title[Classification, inference and segmentation of anomalous diffusion with RNN]{Classification, inference and segmentation of anomalous diffusion with recurrent neural networks
}

\author{Aykut Argun}
\address{Department of Physics, University of Gothenburg, Origov\"agen 6B, SE-41296 Gothenburg, Sweden}
\author{Giovanni Volpe}
\address{Department of Physics, University of Gothenburg, Origov\"agen 6B, SE-41296 Gothenburg, Sweden}
\author{Stefano Bo}
\address{Max Planck Institute for the Physics of Complex Systems, N\"othnitzer Stra\ss e 38, DE-01187 Dresden, Germany}
\ead{stefabo@pks.mpg.de}

\date{\today}

\begin{abstract}
Countless systems in biology, physics, and finance undergo diffusive dynamics.
Many of these systems, including biomolecules inside cells, active matter systems and foraging animals, exhibit anomalous dynamics where the growth of the mean squared displacement with time follows a power law with an exponent that deviates from $1$.
When studying time series recording the evolution of these systems,
it is crucial to precisely measure the anomalous exponent and confidently identify the mechanisms responsible for anomalous diffusion.
These tasks can be overwhelmingly difficult when only few short trajectories are available, a situation that is common in the study of non-equilibrium and living systems.
Here, we present a data-driven method to analyze single anomalous diffusion trajectories employing recurrent neural networks, which we name RANDI. We show that our method can successfully infer the anomalous exponent, identify the type of anomalous diffusion process, and segment the trajectories of systems switching between different behaviors.
We benchmark our performance against the state-of-the art techniques for the study of single short trajectories that participated in the Anomalous Diffusion (AnDi) Challenge.
Our method proved to be the most versatile method, being the only one to consistently rank in the top $3$ for all tasks proposed in the AnDi Challenge.
 
\end{abstract}
\maketitle

\section{Introduction}
Many physical and biological systems feature anomalous diffusion dynamics.
Prime examples are the motion of microscopic particles in a crowded subcellular environment, the active  dynamics of biomolecules inside cells, the displacement of tracers in turbulent flows, and the evolution of financial time series \cite{Klafter2005,Barkai2012, Hofling2013, Metzler2014}.
While normal diffusion is characterized by a linear growth of the mean squared displacement (MSD) with time, anomalous diffusion display a non-linear, power-law growth.
For a system described by a variable $\mathbf{X}(t)$, its MSD is
\begin{equation}\label{eq:def}
    {\rm MSD}(t) 
    = {\rm E} \left[
       \big|\mathbf{X}(t) -\mathbf{X}(0)\big|^2
    \right]
    \propto t^\alpha,
\end{equation}
where $\alpha$ is the anomalous diffusion exponent 
and ${\rm E}[\cdot]$ denotes the ensemble average. 
The exponent $\alpha$ discriminates between normal diffusion ($\alpha = 1$) and anomalous diffusion ($0 < \alpha < 1$ for sub-diffusion, and $1 < \alpha < 2$ for super-diffusion).

Different processes can give rise to anomalous diffusion and several models have been developed to describe them \cite{Metzler2014}.
Correctly identifying the model describing the underlying process and estimating the anomalous exponent provide key insights on the microscopic dynamics and the macroscopic behavior of the systems under consideration. 
It is therefore important to devise reliable methods to extract this information from experimental data.
When the experimental conditions allow to obtain a large ensemble of measurements under identical conditions, it is possible to directly obtain the anomalous exponent by fitting the empirical MSD \cite{Golding2006,Bronstein2009,Weber2010,Jeon2013,Caspi2000} or by using other techniques or statistical observables \cite{Tejedor2010,Burnecki2015,Meroz2015,Makarava2011, Hinsen2016,Krapf2019,Weron2019}. Under these conditions, the model describing the underlying process can also be identified \cite{Meroz2015,Magdziarz2009,Burnecki2012,Regner2013,Thapa2018,Aghion2021}.

\begin{table*}
{\footnotesize
    \begin{tabular}{p{0.23\textwidth}p{0.23\textwidth}p{0.23\textwidth}p{0.23\textwidth}}
        \begin{center}
            {\bf Task} 
        \end{center}
    & 
        \begin{center}
            {\bf RANDI Architecture}
        \end{center}
    & 
        \begin{center}
            {\bf Output}
        \end{center}
    & 
        \begin{center}
            {\bf Loss}
        \end{center}
    \\
    \hline
        \begin{center}
        Classification of the anomalous diffusion model
        \end{center}
    &
        \begin{center}
        2 LSTM layers (250,50) \\
        Dense layer (20) \\
        Output Layer (5)
        \end{center}
    &
        \begin{center}
        Probability of the input belonging to each model.  
        \end{center}
    &
        \begin{center}
        Categorical cross entropy
        \end{center}
    \\
    \hline
        \begin{center}
        Inference of the anomalous diffusion exponent
        \end{center}
    &
        \begin{center}
        2 LSTM layers (250,50) \\
        Output Layer (1)
        \end{center}
    &
        \begin{center}
        Anomalous diffusion exponent.
        \end{center}
    &
        \begin{center}
        Mean squared error
        \end{center}
    \\
    \hline
        \begin{center}
        Classification of the anomalous diffusion segments
        \end{center}
    &
        \begin{center}
        2 LSTM layers (250,50) \\
        Dense layer (20) \\
        Output Layer (5) \\
        \end{center}
    &
        \begin{center}
        Probability of the input belonging to each model. (Two networks are trained for two segments)
        \end{center}
    &
        \begin{center}
        Categorical cross entropy
        \end{center}
    \\
    \hline
        \begin{center}
        Parameter inference of the segmentation of anomalous diffusion trajectories
        \end{center}
    &
        \begin{center}
        2 LSTM layers (250,50) \\
        Dense layer (20) \\
        Output Layer (4)
        \end{center}
    &
        \begin{center}
        Anomalous diffusion exponent of the first and second model, sine and cosine of the normalized distance of the switching time. 
        \end{center}
    &
        \begin{center}
        Mean squared error
        \end{center}
    \\
    \hline
    \end{tabular}
}
\caption{{\bf Information table for different architectures used by RANDI for different anomalous-diffusion-related tasks.} In all architectures, we use the same 2-LTSM-layer structure. If the output layer has more than a single node, we use an additional dense layer of $20$ nodes. We use mean squared error as the loss function for inference tasks and categorical cross entropy for classification tasks.}
\label{table1}
\end{table*}

\begin{figure}[h]
	\begin{center}
		\includegraphics[width=.8\textwidth]{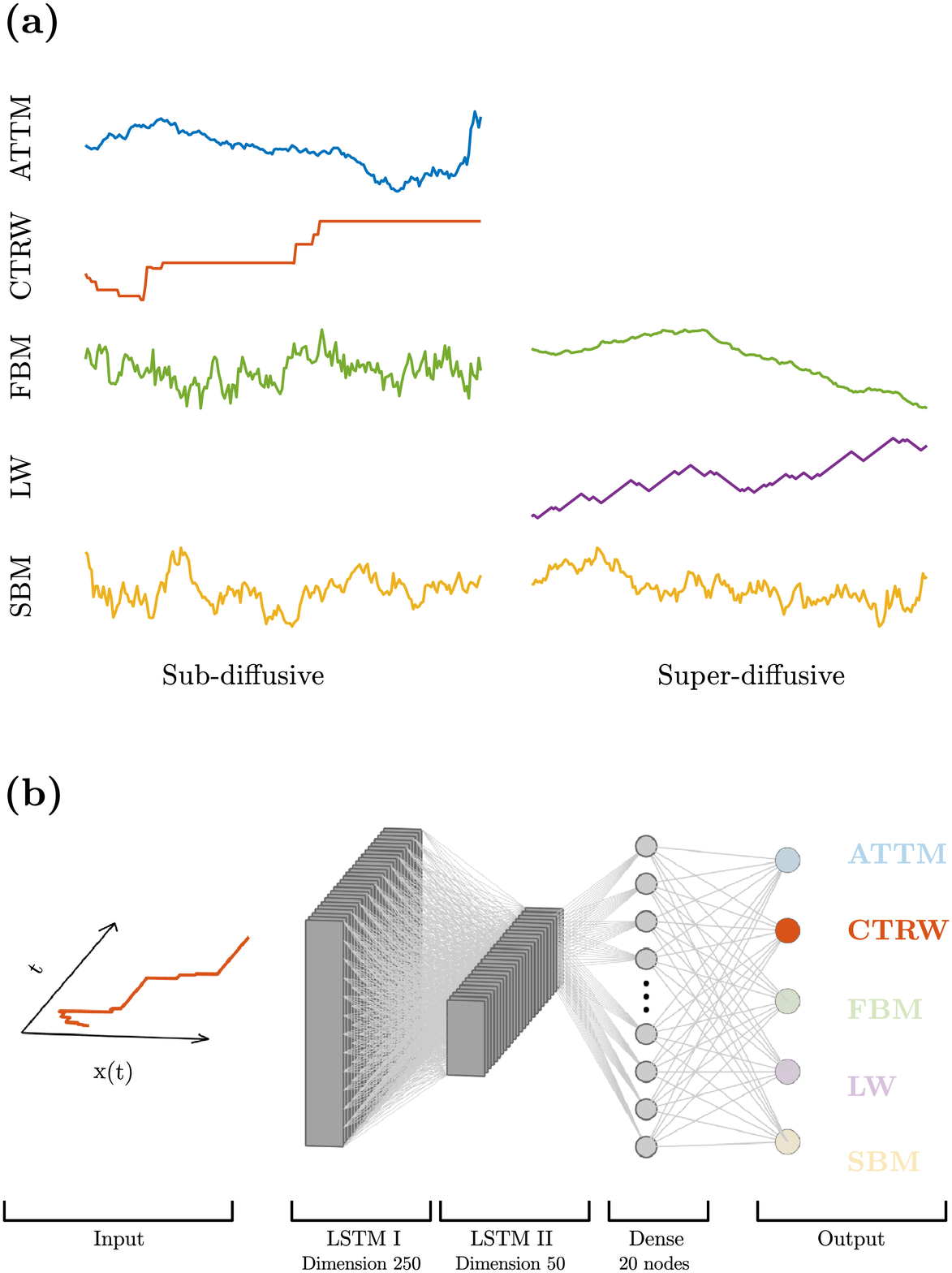}
		\caption{{\bf RANDI architecture to classify the model underlying anomalous diffusion.} 
		{\bf (a)} Sample sub-diffusive (left panels) and super-diffusive (right panels) trajectories for ATTM (blue line), CTRW (orange line), FBM (green lines), LW (purple line), and SBM (yellow lines). The ATTM and CTRW models cannot produce super-diffusive trajectories, the LW model cannot produce sub-diffusive trajectories.
		{\bf (b)} The structure of the recurrent neural network (RNN) we use for identifying the model underlying the anomalous diffusion trajectories. The input trajectory (in this example, CTRW) is fed into the first long short-term memory (LSTM) layer with dimension 250, which is followed by the second LSTM layer with dimension 50. Outputs from the second LSTM layer are then fed into a fully connected dense layer of 20 nodes. Finally, the dense layer is connected to output activation nodes that identify the probability of the input belonging to each class.}
	\end{center}
	\label{fig:fig1}
\end{figure}
However, it is usually difficult to obtain large data sets under tightly  controlled experimental conditions, especially when dealing with non-equilibrium or biological systems \cite{Jeon2010a}.
Then, one often has to work with few, short, non-stationary trajectories  \cite{Weron2019,Meroz2015,Elf2019,Akin2016,Sikora2017c,Weron2017}).
These conditions can be very challenging for the standard approaches and new techniques are usually developed on a case-by-case basis. Additionally, the presence of measurement noise further complicates the task. Indeed, one has to find a fine balance between removing the noise (for instance by some averaging) and avoiding distorting the signal, which is very difficult to achieve algorithmically  \cite{Kepten2013,Burnecki2015,Kepten2015,Lanoiselee2018}. 
It is then natural to consider alternative data-driven approaches, which have established themselves as additional tools to characterize experimental data in physics and biology \cite{Zdeborova2017,Cichos2020, Argun2020}.
Many of these methods are supervised machine learning techniques, where
the basic idea is that, instead of using fixed rules to process the data (as is done in usual algorithmic approaches), the optimal rules to extract information from the data are learnt in a supervised manner by considering large data sets of input data labeled with the sought-after feature.  
For challenging cases, where the limitations of traditional approaches are more evident, data-driven approaches can make a more efficient use of the information contained in the available data. 
For anomalous diffusion, the potential of machine-learning techniques has already been shown by a number of recent studies \cite{Wagner2017,Bo2019, Munoz-Gil2020a,Granik2019,Jamali2020,Janczura2020,Kowalek2019,Loch-Olszewska2020,Han2020,Gentili2021}.

In this article, we present a method based on recurrent neural networks (RNN) \cite{Lipton2015a,Hochreiter1997} to extract anomalous-diffusion information from short individual trajectories, which we name RANDI.
This method uses RNN that have an architecture which is similar to the one we have used for the calibration of force fields and the inference of the anomalous diffusion exponent from single trajectories in  \cite{Bo2019,Argun2020}. 
With the name RANDI, we refer to the whole method, which means the use of different RNN and their suitable combination for each of the tasks we consider.
We focus on three main tasks, which concern the inference of the anomalous diffusion exponent, the classification of the model giving rise to anomalous diffusion, and the segmentation of trajectories that switch between different behaviors. RANDI provides a robust method to perform classification and inference on trajectories generated from different anomalous diffusion models, with different anomalous exponents and subject to varying degrees of measurement noise without a priori knowledge on the model or the noise level. 

\paragraph*{The Anomalous Diffusion Challenge.}
To allow a systematic analysis of the performance of our method and to benchmark it against alternative approaches, we rely on the tasks, models, exponents and noise proposed in the Anomalous Diffusion Challenge (AnDi Challenge \cite{Munoz-Gil2020}), which we briefly describe here.
The numerical methods to generate the trajectories for the training and testing are the ones provided by the AnDi package \cite{Munoz-Gil2020b}.
The five models for anomalous diffusion proposed by AnDi have widely differing features as shown in Fig.~\ref{fig:fig1}(a). These models are:
\begin{itemize}
    \item Continuous time random walks (CTRW): a particle waits in a certain position for a given time, which is randomly chosen, and then takes a step of random length \cite{Scher1975}.
    As in the AnDi challenge, we focus on waiting times following a power-law distribution $\psi(\tau)\sim \tau^{-\sigma}$ and steps with lengths generated by a normal distribution.
    The anomalous diffusion exponent is then given by $\alpha = \sigma-1$.
    We consider $1.05\le \sigma\le2$ for which $0.05 \le \alpha \le 1$, meaning the the dynamics is sub-diffusive and diffusive. This model is non-ergodic.
    
    \item Annealed transient time motion (ATTM) is characterized by the Brownian motion of a particle, with a randomly changing diffusion coefficient \cite{Massignan2014}. As in the AnDi challenge, we focus on the case in which the diffusion coefficient varies in time. The particle diffuses with a given diffusion coefficient for certain time $\tau$ and then switches to a different coefficient. The diffusion coefficients are drawn randomly from a power-law distribution $P(D)\sim D^{\sigma-1}$ for  $D\le1$. 
    Once $D$ is chosen, 
    the duration $\tau$ of an interval with this value of  $D$ is fixed to be  $\tau=D^{-\gamma}$. 
    We consider $\sigma\le \gamma<\sigma+1$, which gives an exponent $\alpha=\sigma/\gamma$,  with sub-diffusive and diffusive dynamics $0.05 \le \alpha \le 1$. The duration of the intervals with constant diffusion coefficient follows a power law $\psi(\tau)\sim \alpha \tau^{-1-\alpha}$ and this model is non-ergodic.

    \item Fractional Brownian Motion (FMB) is characterized by correlated Gaussian increments \cite{Mandelbrot1968}. When the increments are positively (negatively) correlated the model displays super-diffusion with $1 < \alpha < 2$ (sub-diffusion with $0.05\le  \alpha < 1$). This model is ergodic.

    \item L\'evy walks (LW) display irregular waiting times $\tau$ as CTRW but with a step length that is determined by the waiting time \cite{Klafter1994}. The step  $\Delta \mathbf{x}$ is chosen such that the speed $v=|\Delta \mathbf{x}|/\tau$ is constant. We consider a power law distribution of waiting times and diffusive and super-diffusive dynamics ( $1 \le \alpha \le 2$). This model is non-ergodic.

    \item Scaled Brownian motion (SBM) is characterized by the Brownian motion of a particle, with a deterministically varying diffusion coefficient \cite{Lim2002}. We consider the case in which $D(t)=\alpha D t^{\alpha-1}$, and both sub-diffusive and super-diffusive (dynamics $0.05 \le \alpha \le 2$).
    This model is non-ergodic.
  
\end{itemize}
The trajectories generated by these models are then corrupted by adding measurement noise of varying signal-to-noise ratios (SNR). Specifically, we consider Gaussian localization noise of low, medium and high intensity corresponding to 10\% (SNR 10), 50\% (SNR 2) and 100\% (SNR 1) of the standard deviation of the single trajectory displacements, respectively. The AnDi challenge focuses on short trajectories of varying lengths ranging from just $10$ measurement points to $1000$ measurement points.

The AnDi challenge consists of three main tasks: classification of the anomalous diffusion model, inference of the anomalous diffusion exponent, and characterization of a trajectory switching between different regimes (model classification, exponent inference and changepoint localization).
Each task is split into sub-tasks depending on the spatial dimension of the considered trajectories ($1$-, $2$- and $3$-dimensional), giving a total $9$ tasks. 
Notice that considering such a diverse set of models (see Fig.~\ref{fig:fig1}(a)) makes it very difficult to devise a single algorithmic approach that would perform well on all of them. 
Just to give an example, all  considered model except FBM are  non-ergodic so that it is not possible to just extract the exponent from the dependence on the lag-time of time-averaged MSD \cite{Jeon2010a,Massignan2014}.
This motivates the use of data-driven techniques.

RANDI is a refined version of the methods we used to participate in 
the AnDi challenge. Our team in the challenge was named eduN and our method proved to be the most versatile one.
In particular, we were the only team to rank in the top $3$ for all tasks proposed in AnDi and we developed the top-ranked method for classification (in 1 and 2 dimensions) and for the characterization of switching trajectories (in 1 and 3 dimensions).
RANDI manages to reliably extract information in scenarios that are overwhelmingly challenging for traditional algorithmic methods and performs better than all methods participating in the AnDi challenge in 8 tasks out of 9. 
We make RANDI freely available as a Python package \cite{RANDI_git}.
\section{Neural network architecture and training procedure}\label{sec:architecture}

We tackle the three different sets of tasks with three different types of neural networks that all share some core features. 
All networks employ two layers of recurrent neural networks called long short-term memory (LSTM) \cite{Hochreiter1997}, with dimensions 250 and 50, respectively (as shown in Fig.~\ref{fig:fig1}(b) and in Table~\ref{table1}).
For classification tasks (i.e., prediction of the anomalous diffusion model class), the last output of the second LSTM layer is followed by a dense layer (20 nodes) and an output layer of 5 nodes (with softmax activation, each associated with the probability of belonging to a specific model), as shown in Fig.~\ref{fig:fig1}(b). For inference tasks (i.e., prediction of the anomalous diffusion exponent), the last output of the second layer is connected directly to an output layer. The detailed description of the different RNN architectures is shown in Table~\ref{table1}.  
The use of multiple LSTM layers allows the neural network to reveal both short and long term correlations and prevents the training from becoming unstable \cite{Lipton2015}.
In addition, recent studies have shown that the LSTM networks perform very well on stochastic trajectories \cite{Bo2019,Argun2020} and can work on trajectories of different lengths without the need of adapting the trajectory to any fixed length. This avoids the need for padding (which might alter correlations) or slicing (which reduces the number of available measured points) of the input trajectory --- a significant advantage with respect to other artificial neural networks, such as dense and convolutional neural networks. We decided the number of layers and their dimension in such a way that we reach a sufficient level of complexity without incurring in the diverging gradient problem \cite{Bottou2018}.

For neural-network-based methods (as for every supervised machine-learning technique), it is of great importance to accurately choose the data sets on which the method is trained.
We train our method on data sets with the same structure as those provided by the AnDi challenge \cite{Munoz-Gil2020a,Munoz-Gil2020b}. This means that we sample uniformly the anomalous diffusion exponents and consider three levels of measurement noise (each sampled with equal probability).  
We also generate roughly the same number of trajectories for each of the five models of anomalous diffusion. This ensures robustness of the method and high performance on data with the same kind of heterogeneous features. 

As we will discuss in details below, the most effective ways of extracting information from single trajectories depend on the trajectory length.
For this reason, we trained different networks (with the same architecture) on different data sets each containing trajectories of the same length.
Each network is trained on  $3\times 10^6$ simulated  trajectories, which we 
split in 30 smaller sets each containing $10^5$ trajectories. These data sets are progressively presented to the RNN to prevent overfitting. We use the first data set (split in batches of size $32$) to train for $5$ epochs. We then switch to another data set (now split in batches of size $128$) and train for $4$ epochs. We repeat this procedure for $3$ other data sets. 
We then use other $5$ data sets split into batches of size $512$ each considered for $3$ epochs and finally use $20$ data sets (split in batches of size $2048$) for $2$ epochs each\footnote{With the exception that for long two- and three-dimensional  trajectories, for memory reasons maximum batch size was limited to 1024.}. We use 20 \% recurrent dropout in both layers for all trained networks in order to prevent overtraining \cite{Gal2016}. 
The networks for inference tasks are trained with mean squared error as loss function and the ones for classification tasks with  categorical crossentropy. We used the Adam optimizer for all tasks \cite{Kingma2015}. 
Both for training and testing, we apply a minimal pre-processing to the input trajectories. The input to the network is given by the increments of each trajectory ($\Delta x_i = x_{i+1}-x_i$),  normalized such that for each trajectory we have that the mean of the increments  $\overline{\Delta x}=0$ and their variance $\sigma^2(\Delta x)=1$. 
To make the training more efficient, we group the input increments in blocks of a chosen dimension that is chosen specifically for each network. For example, for the inference of one-dimensional trajectories of length $125$, the $124$ increments are split such that the input is given by $31$ blocks of dimension $4$, $b_j= [\Delta x_{4j}, \Delta x_{4j+1},\Delta x_{4j+2}, \Delta x_{4j+3}]$  with $j=0, \ldots 30$.
For trajectories that have higher dimension, we reshape the input trajectory such that each block contains measurements in all the spatial dimensions of the systems within its time frame (e.g., for a two-dimensional system $[\Delta x_0, \Delta y_0,\Delta x_1, \Delta y_1, ...]$). 
The trajectories need to contain a number of measurement points that is a multiple of the block size. This means that trajectories might need to be slightly cut to fit this requirement. This procedure is harmless for long or high dimensional trajectories but it can start playing a role for shorter ones. For this reason, we adapt the block sizes for different trajectory lengths and dimension. 
For one-dimensional classification, we use blocks of size $2$ for the network trained on trajectories of length  $25$ and $75$ and block size $4$ for the remaining ones. For two-(three-) dimensional classification, we use blocks of size $4$ ($6$) for the networks trained on trajectories of length  $25$ and $65$ and block size $8$ ($12$) for the other ones.
For one-dimensional inference, we use blocks of size $1$ for the networks trained on trajectories of length  $25$ and $50$, block size $2$ for trajectories of length $75$ and block size $4$ for the remaining ones.
For two-dimensional inference, we use blocks of size $4$ for trajectories of length $25$ and blocks of  size $8$ for the other ones.
For three-dimensional, inference we use blocks of size $6$ for trajectories of length $25$ and $65$ and blocks of  size $12$ for the other ones.
For the segmentation tasks we consider trajectories of fixed length containing $200$ points. After normalizing the input trajectories we add a $0$ at the end and reshape them using blocks of size $2$, $4$ and $6$ for data of dimension $1$, $2$ and $3$, respectively.

\section{Classifying different models of anomalous diffusion}

The first task we consider is the classification of the model giving rise to the anomalous diffusion dynamics.
To assess our performance, we focus on the accuracy of our classification, defined as the percentage of correctly classified trajectories.

\begin{figure}[h t]
	\begin{center}
		\includegraphics[width=\textwidth]{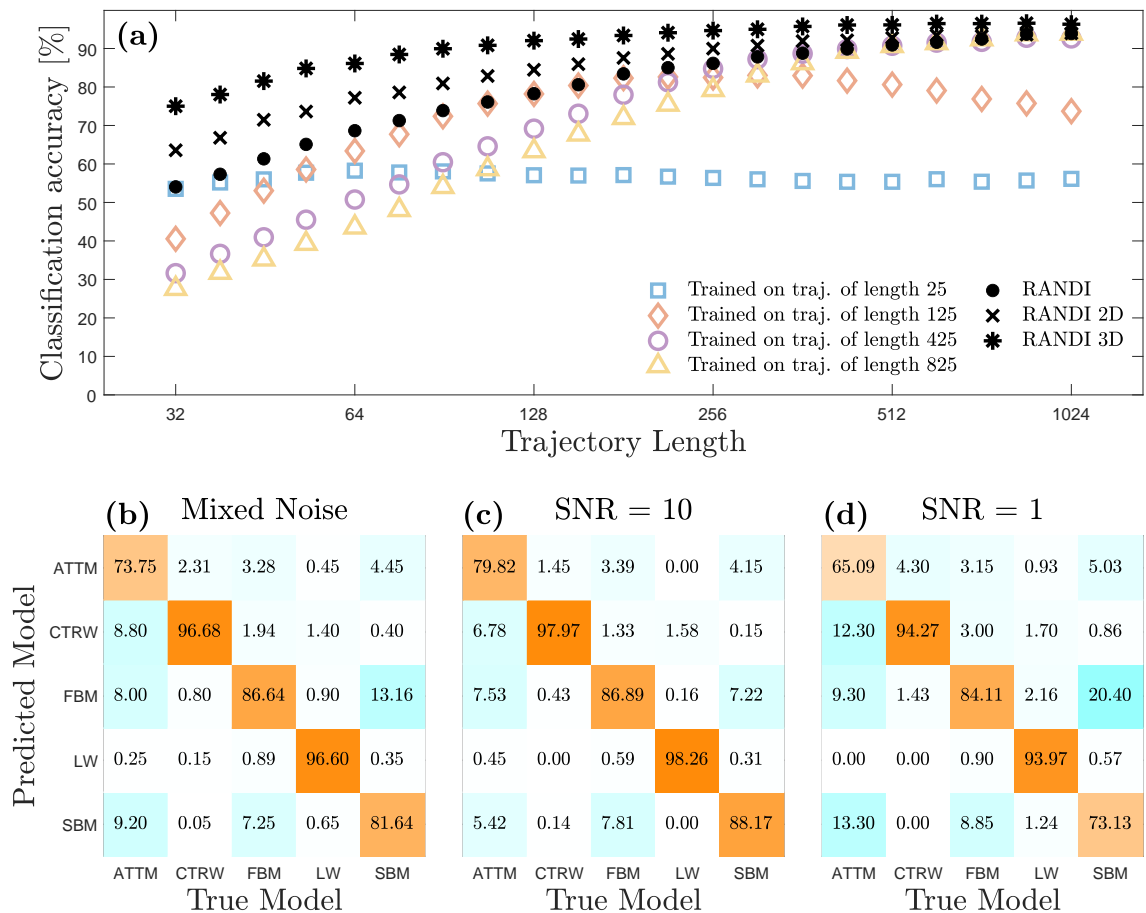}
	\end{center}
	\caption{{\bf Classification of the model underlying anomalous diffusion.} 
	{\bf (a)} Classification accuracy for the anomalous diffusion models from single trajectories of different length. The performance of RNNs that are trained with trajectories of length 25 (blue squares), 125 (orange diamonds), 425 (purple circles) and 825 (gold triangles) are shown as a function of input length of the validation data for 1D data. 
	The performance of RANDI obtained combining the predictions from different networks is shown as black dots for 1D data. 
	The performance for combined networks trained on 2D (3D) data tested on 2D (3D) data is shown as black crosses (black asterisks).
	{\bf (b-d)} Confusion matrices for the classification of the different models of anomalous diffusion applied to the 1D scoring data set of AnDi {\bf (b)} for the mixed noise (trajectories with SNR $10$, $2$ and $1$), {\bf (c)} for only the data with low noise (SNR $10$), and {\bf (d)} for only the data with high noise (SNR $1$).}
	\label{fig:fig2}
\end{figure}

\paragraph*{Dependence of classification performance on trajectory length.}
How to efficiently 
extract information from short trajectories and filter out noise depends on the amount of available data, i.e., on the length of the trajectory \cite{Kepten2013,Burnecki2015,Kepten2015,Lanoiselee2018}. We therefore train different networks for different trajectory lengths.
In principle, training networks for as many different trajectory lengths as possible would yield the best performances. However, this would be very time consuming and overall  unnecessary thanks to the robustness of the networks.
The open symbols in Fig.~\ref{fig:fig2}(a) show how the performance of networks trained on a set of trajectories of a specific length varies with the length of the trajectories on which they are tested. The plot shows that each network excels for trajectories of length similar to its training set and that it gets outperformed by other networks for very different trajectory lengths, even though we notice a remarkable robustness. 
This feature suggests to employ, for each test trajectory length, a  combination of the networks trained on the nearest trajectory length.
Since networks trained on short trajectories are more sensitive to the length of the trajectories on which they are applied we choose to train more networks for short trajectories. In total we train six networks. Each network is trained on a set of trajectories of fixed length  $L_1 = 25$, $L_2=65$, $L_3 = 125$, $L_4 = 225$, $L_5 = 425$ and $L_6 = 825$.
 When classifying a trajectory of length $l$ we apply the two networks trained on trajectories just longer and just shorter than $l$\footnote{Trajectories that are shorter (longer) than the shortest (longest) trajectories considered for the training are treated differently. For these trajectories we simply use the network trained on the shortest (longest) trajectories.}. We weight the contribution of each network according to the distance between the length of the trajectories on which it was trained and the length of the analyzed trajectory $l$. Specifically, denoting by $\mathbf{o}_i$ the output of the network trained on trajectories of length $L_i$, 
 the output of RANDI is given by
 \begin{equation}
    \mathbf{o} = d \mathbf{o}_i+(1-d)\mathbf{o}_{i+1}
    \label{eq:comb}
 \end{equation}
with $d=(L_{i+1}-l)/(L_{i+1}-L_i)$ and where $i$ is chosen such that $L_i$ is the largest length fulfilling $L_i\le l$.
This combination achieves an overall accuracy on the AnDi data set of $87.05\%$, which was the highest score in the challenge out of $14$ submitted methods.
 The black dots in Fig.~\ref{fig:fig2}(a) show that, as expected, the accuracy increases with trajectory length starting from about $54\%$ for trajectories containing only $32$ measurement points and reaching a value of about $94\%$
for trajectories of length $1024$.

\paragraph*{Dependence of classification performance on  trajectory model.}
The confusion matrix (Fig.~\ref{fig:fig2}(b)) shows that the classification performance strongly depends on the model classes.
CTRW and LW can be identified with very high confidence.
This may be due to the fact that these two models differ strongly from the other models and are characterized by long interval of times during which the particle is trapped (CTRW) or moves with a roughly
constant velocity (LW) so that, during these intervals, their displacement increments can be highly correlated (see sample trajectories in Fig~\ref{fig:fig1}(a)). 
In contrast, FBM and SBM are often hard to tell apart from each other. ATTM is the hardest model to identify and often gets mistaken for another of the sub-diffusive models (CTRW, FBM or SBM). This feature is related to the fact the certain short trajectories of ATTM are actually indistinguishable from normal diffusion, as we shall discuss in more detail in section~\ref{sec:task1}. Normal diffusion trajectories are especially difficult to classify. In particular, for $\alpha=1$, trajectories generated by FBM and SMB are just simple diffusion processes with a constant diffusion coefficient and are therefore intrinsically indistinguishable.

\paragraph*{Dependence of classification performance on noise level.}
Unsurprisingly, the presence of measurement noise complicates the classification. As mentioned before, we train our networks on data that contain different levels of measurement noise, having approximately the same number of trajectories with low, medium and high levels of noise. 
This procedure makes our networks fairly robust against noise. However, noisier trajectories remain harder to classify.
For low levels of noise with localization noise corresponding to 10\% of the standard deviation of the displacements, {\it i.e.} signal-to-noise ratio (SNR) 10, RANDI achieves an overall accuracy of $90.2\%$ in classification. The confusion matrix for low levels of noise (SNR 10) is shown in Fig.~\ref{fig:fig2}(c). For intermediate noise levels (SNR 2), 
the performance declines only marginally and the method has an accuracy of $89.3\%$, even thought the noise increased by a factor of 5. We have to push to high noise levels (SNR 1) to observe a noticeable decrease in the performance with an accuracy of $81.7\%$. The confusion matrix for high levels of noise (SNR 1) is shown in Fig.~\ref{fig:fig2}(d).
Noise does not disrupt classification homogeneously, as some models remain easy to classify as evident from the confusion matrices of Figs.~\ref{fig:fig2}(b-d).
The models that suffer the most are those for which identification is harder even in the presence of low noise levels (ATTM and SBM) with the exception of FBM, which continues to be correctly identified.
With high noise levels, only $65\%$ of the trajectories generated following ATTM are correctly identified as it gets more commonly mistaken for CTRW ($12\%$ of the times) and for SBM ($13\%$ of the times). 
SBM also becomes harder to identify and is mistaken for FMB in $20\%$ of the cases while this happens only in $7\%$ of the cases for low noise.

\paragraph*{Higher dimensions.}
There are different ways to generalize the one-dimensional anomalous diffusion processes we are considering to higher dimensions.
We follow the procedure proposed in the AnDi challenge.
For FBM and SBM, the higher dimensional dynamics are simply given by independent one-dimensional realizations in each of the spatial dimensions.
For LW,  
we generate the length of the displacements in accordance to a one-dimensional LW and then randomly choose the direction of the displacement.
For CTRW and ATTM, in two dimensions we generate a single sequence of waiting times and times with constant diffusion coefficients, respectively. The displacements are generated independently along the two spatial dimensions.
In three dimensions, CTRW and ATTM are generated similarly to LW, i.e., the step length is obtained as in a one-dimensional CTRW or ATTM model and the step direction is chosen randomly. 
The fact that some models are not a simple composition of independent one-dimensional realizations along the different dimensions prevents a straightforward application of the networks trained on one-dimensional trajectories.
We thus train new networks on two-dimensional and three-dimensional data. 
Every trajectory contains two or three times the number of data points compared to the one-dimensional case, making the training more expensive.
We therefore train only five networks on trajectories of length $L_i \in \{25, 65, 125, 225, 425\}$.
Their performances as a function of trajectory length is shown in supplementary figure~\ref{fig:suppfig}. 
Combining their predictions gives an  accuracy of $89.16\%$ for the two-dimensional trajectories of the AnDi data set. This performance is higher than the best result submitted to the challenge, where we had used only four networks and a different reshaping of the input trajectories. 
For three-dimensional data, 
we obtain an accuracy of $93.71\%$, which is higher than the best performance submitted to the challenge.
In our submission to the challenge, due to time constrains, we had not trained three-dimensional networks. 
The performances of RANDI as a function of trajectory length are shown in Fig.~\ref{fig:fig2} where the crosses and asterisk symbols refer to two and three-dimensional data, respectively.
\section{Inferring the anomalous diffusion exponent}\label{sec:task1}

The second task we consider is the inference of the anomalous diffusion exponent.
In the idealized case in which one has access to many long trajectories without measurement noise, this task can be readily achieved, for instance by measuring the (ensemble-averaged) mean squared displacement and observing its power-law dependence on the lag-time.
This becomes harder if we do not have access to multiple trajectories since most of the models of anomalous diffusion we are considering 
are not ergodic such that the time average over a single trajectory differs from the ensemble average over many trajectories. 
Inference for single trajectories is therefore difficult and increasingly so as the length of the trajectories gets shorter. 
To complicate the analysis further, not all standard techniques to infer the anomalous diffusion exponent from single trajectories can be directly used without knowing the model of 
anomalous diffusion under consideration (see \cite{Aghion2021,Aghion2021a} for exceptions). 
Measurement noise also plays an important role, often requiring a complex pre-processing of the data, which might also depend on the underlying anomalous diffusion model.
This highlights the importance of obtaining information about the underlying model before applying a standard technique. A possibility would be to first determine the underlying model (as discussed in the previous section) or some of its properties (e.g., ergodicity) and then choose what technique to use.
However, this kind of approach
has the shortcoming that errors rapidly propagate and that inaccuracies in identifying the model lead to the wrong choice of the inference method, which can have dramatic negative consequences on the inference performance.
Training a neural network to tackle the problem in one go can mitigate error propagation as the network can, in principle, implicitly compensate for uncertainties about the model. 
This is therefore our approach of choice.

\begin{figure}[h!]
	\begin{center}
	\includegraphics[width=.72\textwidth]{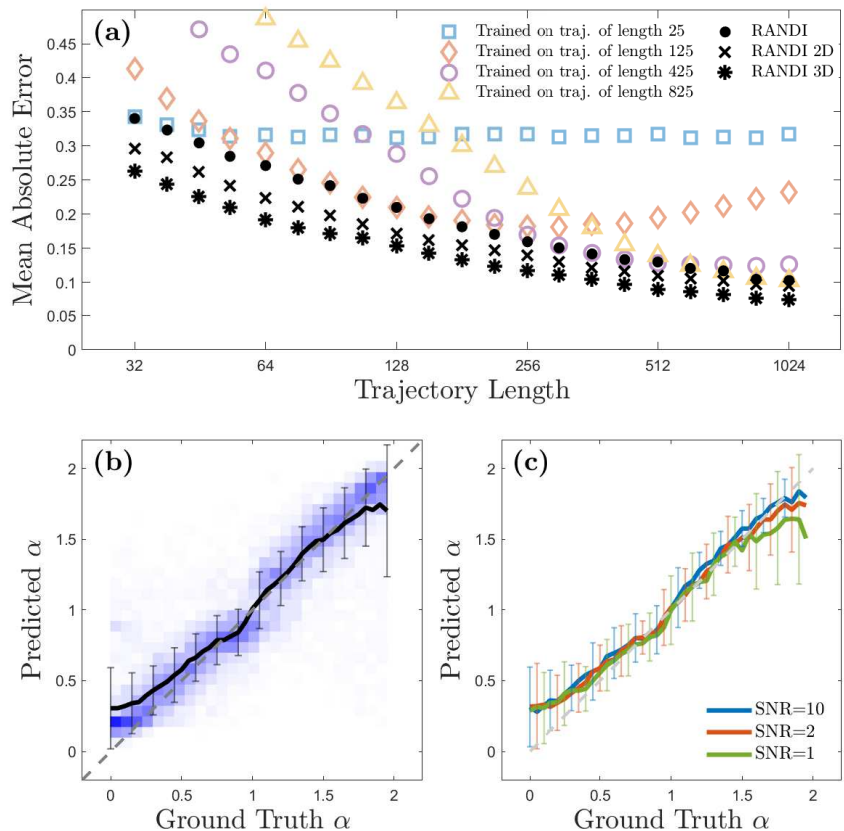}
	\end{center}
	\caption{{\bf Inference of anomalous diffusion exponent as function of trajectory length and noise.} 
	{\bf (a)} Mean absolute error when measuring the anomalous diffusion exponent ($\alpha$) of single trajectories as a function of length. The performance of RNNs trained with 1D trajectories of length 25 (blue squares), 125 (orange diamonds), 425 (purple circles) and 825 (gold triangles) are shown as a function of input length of the validation data.
	The performance of RANDI, obtained combining the predictions from different networks is shown as black dots for 1D data. 
	The performance for combined networks trained on 2D (3D) data tested on 2D (3D) data is shown as black crosses (black asterisks).
	{\bf (b)} Distribution (blue density plot) of the predictions by RANDI as a function of the ground truth anomalous diffusion exponent ($\alpha$). The black line represents the mean of the predictions and the error bars represent the standard deviation. 
	{\bf (c)} Mean predicted $\alpha$ as a function of the ground truth for low noise (blue line, SNR=10), medium noise (red line, SNR=2) and high noise (green line, SNR=1) levels. 
	Error bars represent the standard deviation.  
		}
	\label{fig:fig3}
\end{figure}

\paragraph*{Dependence of the exponent inference on the trajectory length.}
As we have already seen for the classification, the most effective ways to deal with the presence of measurement noise depend on the trajectory length~
\cite{Kepten2013,Burnecki2015,Kepten2015,Lanoiselee2018}. Therefore, also in this case, we train different networks for different trajectory lengths to optimize performance.
In Fig.~\ref{fig:fig3}(a) the open symbols show the performance of RNNs trained on trajectories of a fixed length when tested on trajectories of different lengths. As for the classification shown in Fig.~\ref{fig:fig2}(a), we see that the RNNs perform well on trajectories of length similar to the one they have been trained on but get outperformed by the other nets for different trajectory lengths. We see that
inference is more sensitive than classification to trajectory length, suggesting the need to use more networks. We choose to train $14$ networks each on a data set containing trajectories of the same length, respectively $L_i \in \{25, 50, 65, 75, 125, 165, 225, 325, 425, 525, 625, 725, 825,  925\}$.
As for classification, to make a prediction on a trajectory of a certain length, we  combine the predictions made by the two networks trained on the closest lengths [see Eq. (\ref{eq:comb})].
Combining the predictions, we achieve a Mean Absolute Error (MAE) for RANDI on the AnDi challenge data set of $0.1558$, which is the third best score with a MAE that is just 2.4\% higher than the one of the best performing method.
As expected, the inference improves for longer trajectories,  with a MAE that improves from $0.34$ for trajectories containing only $32$ points to  $0.10$ for trajectories containing $1024$ points.
Fig.~\ref{fig:fig3}(b) shows that the method is very accurate for exponents in the range $0.5$ to $1.5$ but starts to systematically deviate close to the edges of the range where it also becomes less consistent (higher variance). This is a common feature of supervised inference close to the edges of the training set but is also related to the specific features of some of the models we are considering, as we will discuss later.

\paragraph*{Dependence of the exponent inference on the noise.}
Fig.~\ref{fig:fig3}(c) shows that RANDI is fairly robust to the presence of measurement noise, especially for sub-diffusive dynamics, both in terms of bias and variance. Super-diffusive dynamics are more sensitive to the presence of noise since for FBM and LW they are characterized by a higher correlation of increments. Noise can lower this correlation and lead to a systematic underestimation of the anomalous diffusion exponent for very persistent cases close to the ballistic limit. In addition to this bias, also the variance of the estimation of strongly super-diffusive trajectories is increased by measurement noise.

\begin{figure}[h]
	\begin{center}
	\includegraphics[width=.95\textwidth]{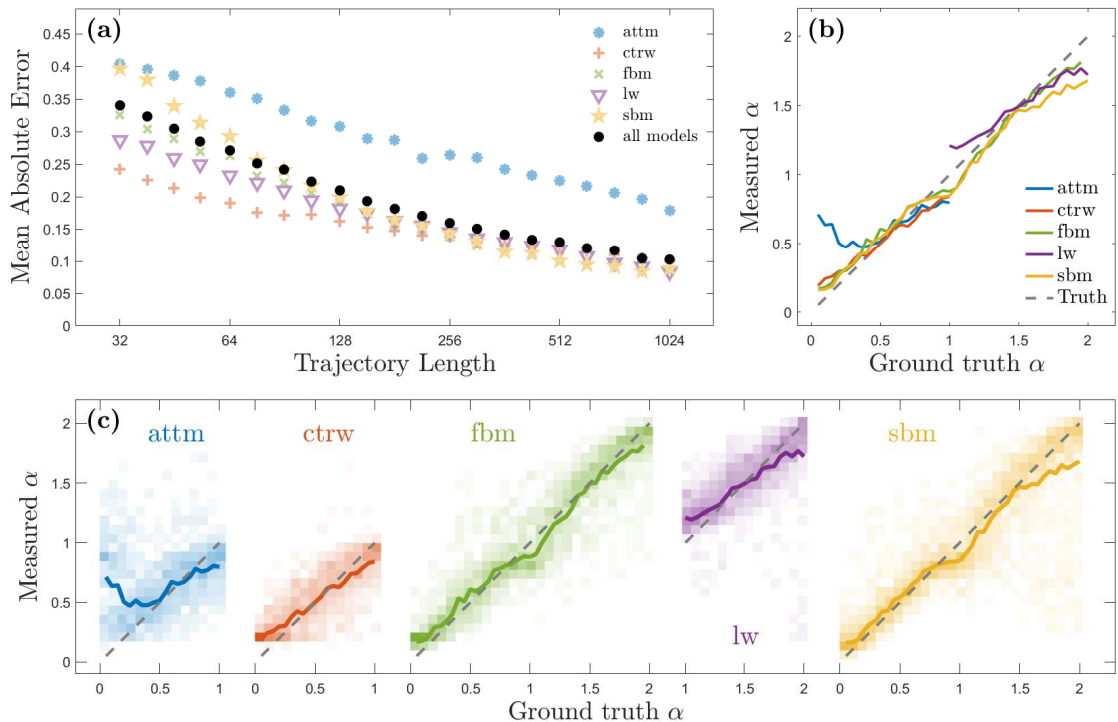}
	\end{center}
	\caption{{\bf Inference of anomalous diffusion exponent as function of diffusion model.} 
	{\bf (a)} Observation of output mean absolute error when measuring the anomalous diffusion exponent ($\alpha$) of single trajectories of different models. Mean absolute error in predicting $\alpha$ as a function of the trajectory length for ATTM (blue dots), CTRW (orange pluses), FBM (green crosses), LW (purple triangles) and SBM (gold stars). The average results for input data set including all the models is shown as black dots. 
	{\bf (b)} Mean predicted $\alpha$ as a function of the ground truth for ATTM (blue lines), CTRW (orange lines), FBM (green lines), LW (purple lines) and SBM (gold lines). 
	{\bf (c)} Distributions of the predictions by the recurrent neural networks as a function of the ground truth diffusion exponent ($\alpha$) for ATTM (blue histogram), CTRW (orange histogram), FBM (green histogram), LW (purple histogram) and SBM (gold histogram). The lines represent the averages and they are identical to the lines in ({\bf b}).  
		}
	\label{fig:fig4}
\end{figure}

\paragraph*{Dependence of exponent inference on the diffusion model.}
As shown in Fig.~\ref{fig:fig4}, the inference performance depends on the anomalous diffusion model. ATTM is harder to infer and the errors are dominated by a systematic tendency towards mistaking strongly sub-diffusive trajectories as normally diffusing ones. This issue is not due to the method of inference, but it is rather an unavoidable feature of the ATTM trajectories. In fact, these trajectories are characterized by switches of the diffusion constant that take place at random times, which follow a power-law decay $\psi(\tau)\sim \alpha \tau^{-1-\alpha}$. For smaller and smaller $\alpha$, it becomes increasingly likely that the duration of an interval with the same diffusion coefficient is larger than the trajectory length ($\tau>T$). The whole trajectory is then characterized by a single diffusion coefficient and hence indistinguishable from ordinary Brownian motion with $\alpha=1$. For instance, for $\alpha=0.1$, half of the time increments are longer than $1000$ time steps, meaning that most trajectories are indistinguishable from ordinary Brownian motion. 
For moderate sub-diffusion $\alpha>0.5$, trajectories with a constant diffusion coefficient become less likely and the method recovers excellent performances with unbiased predictions.
The presence in the training set of ATTM trajectories that are too short to be distinguished from ordinary diffusion but are labeled as having a small anomalous exponents is probably responsible for the underestimation of the exponents for sub-diffusive processes that are almost diffusive $\alpha\simeq 1$.

\paragraph*{Higher-dimensional data.}
As done for the classification task, new RNNs can be trained for two-dimensional trajectories as well.
Since training for higher dimensions is more computationally expensive and the networks seem robust with respect to the trajectories lengths, we train only $9$ networks on trajectories of length $25$, $65$, $125$, $225$, $325$, $425$, $525$, $725$ and $925$.
Their performances as a function of trajectory length is shown in supplementary figure~\ref{fig:suppfig}. 
Having access to more data, we expect an increase in performance and indeed for two-dimensional trajectories we achieve an overall MAE of $0.1345$, which is better than the best score submitted to the AnDi Challenge. 
For our submission to the AnDi challenge we did not have time to train these many networks on two-dimensional data and we had therefore obtained a lower score. The black crosses in Fig.~\ref{fig:fig3}(a) reporting the MAE as a function of trajectory length show how this increase in performance is more marked for short trajectories. 
For three-dimensional data we train networks on trajectories of length $25$, $65$, $125$, $225$, $525$ and $925$.
This approach achieves a MAE error of $0.1109$, which is better than the best score submitted to the AnDi challenge. In our submission to the AnDi challenge, due to time constraints we had not trained networks on three-dimensional data.
The performance on three-dimensional data as a function of trajectory length is shown as asterisk symbols in Fig.~\ref{fig:fig3}(a).
\paragraph*{Specialist networks outperform generalist networks.}
The RNN trained for all different models and on noisy data 
manages to robustly infer the exponent for the general and difficult settings when the underlying model and the measurement noise levels are not known.
The task of inferring the exponent for a specific model is a much simpler one and a network trained only on that specific model can yield far better performances. This becomes even more evident if we further simplify the task by considering trajectories that are not corrupted by measurement noise.
To provide a rough estimate of the importance of knowing for certain the model and the noise level we consider the unfair comparison between the generalist RNNs trained to deal with trajectories that could be generated by five different models of anomalous diffusion and corrupted by measurement noise and the specialist RNNs trained on FBM only without measurement noise \cite{Bo2019}.
We find that, tested on their speciality, the specialist RNNs can achieve a MAE that is roughly half the one of the generalist.
As expected, the specialist RNNs is unable to provide accurate predictions when applied to trajectories generated with different models.
Furthermore, the specialist RNN, having being trained on trajectories without measurements noise, is also less robust to measurement noise.

\paragraph*{Educated networks do not outperform naive networks.}
The significant improvement in performance observed when the model is known a priori highlights the importance of the underlying model for the ability to correctly infer the anomalous exponent. 
We might thus speculate that a technique that is able to correctly classify the model should also have the potential for an accurate inference and that, conversely, ignoring the underlying model might cripple the inference performances.
With the aim of guiding the learning process to focus on the kind of model we train some networks that, in addition to the anomalous diffusion exponent are asked to extract information concerning the anomalous diffusion model.
Specifically, we train the networks to estimate the Noah, Moses and Joseph exponents, which contain specific information about the model of anomalous diffusion~\cite{Chen2017,Meyer2018}.
The Noah exponent is related to the presence of fat tails in 
the increments (typical, for example of Levy walks), the Moses exponent to non-stationarity of the increments (as in the case of ATTM and SBM), and the Joseph one to time correlations in increments (e.g. FBM) \cite{Mandelbrot1968,Chen2017,Meyer2018,Aghion2021}. 
The sum of the exponents determines the anomalous diffusion exponent \cite{Aghion2021}.
These exponents were used for classification of anomalous diffusion trajectories in \cite{Aghion2021,Aghion2021a}.
We observe that adding these exponents to the inference task results in a faster training of the networks, but does not significantly improve the performance, which remains roughly unchanged. 

\section{Segmenting trajectories switching between different anomalous diffusion regimes}
We apply RANDI to the study of trajectories that switch between different anomalous diffusion dynamics. These switches occur at a random time and entail a change in the anomalous diffusion model and/or the exponent. The task includes the prediction of the switching time as well as the model classification and exponent inference for each segment. An example trajectory for a segmentation task is shown in Fig.~\ref{fig:fig5}(a). For RANDI, we train two different types of RNNs. One type of RNN is devoted to the joint inference of the anomalous diffusion exponents of the two segments and the switching time. The other type of RNN classifies the model and we train one RNN for classifying the first segment and one for the second segment, as discussed in detail in section \ref{sec:task3_train}.
\begin{figure}[h]
	\begin{center}
		\includegraphics[width=\textwidth]{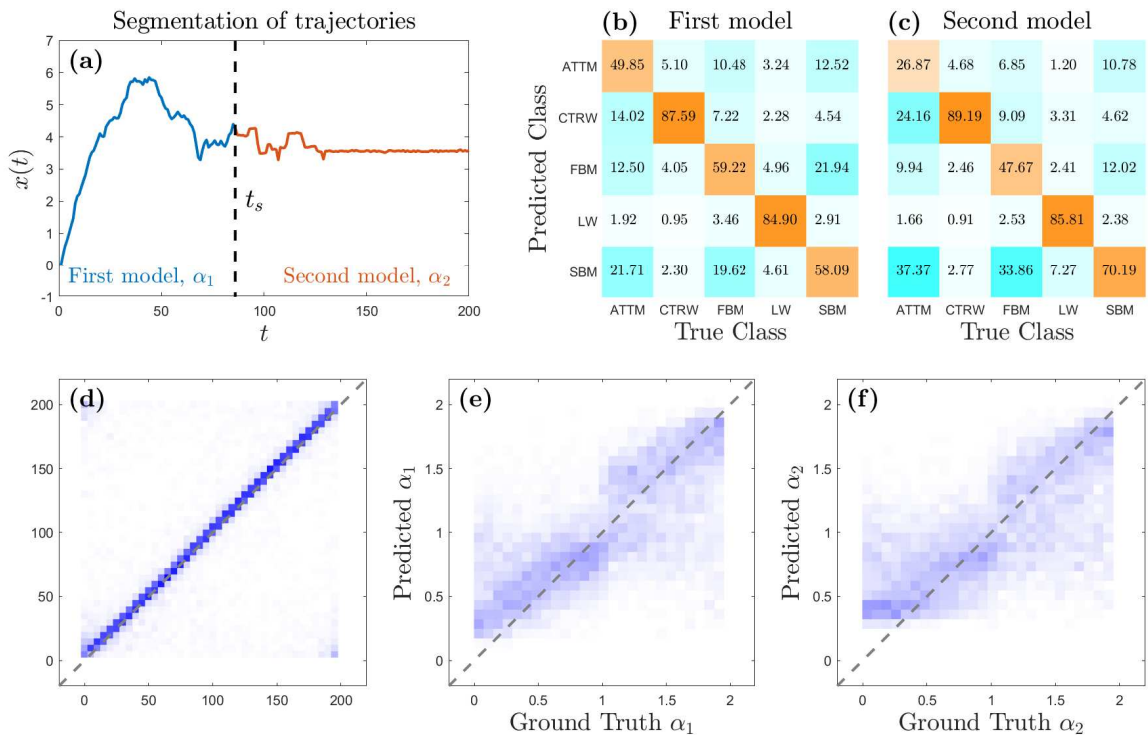}
		\caption{{\bf Segmentation of trajectories switching between different anomalous diffusion regimes.}
		{\bf (a)} Sample input trajectory for the segmentation task. The trajectory starts with the first model and exponent $\alpha_1$(FBM, $\alpha_1$=1.65, blue line), switching at a random time $t_s$ (black dashed line) to the second model and exponent $\alpha_2$ (CTRW, $\alpha_2$=0.8, red line). We predict the first and the second model, $\alpha_1$ and $\alpha_2$, as well as the switching time $t_s$. 
		{\bf (b-c)} Confusion matrices for predicting {\bf (b)} the first and {\bf (c)} the second model. 
		{\bf (d-f)} Distributions of the RANDI predictions as a function of {\bf (d)} the ground truth switching time ($t_s$), {\bf (e)} the ground-truth first exponent ($\alpha_1$), and {\bf (f)} ground-truth second exponent ($\alpha_2$).}
	\end{center}
	\label{fig:fig5}
\end{figure}

In the AnDi challenge this tasks focused on short trajectories of length $200$.
In addition to the difficulty of dealing with such short trajectories,
the presence of the switch greatly increases the complexity of this task.
Indeed, not properly recognizing the switch leads to the confusing task of inferring the exponent (and  classifying the model) from a trajectory that results from the combination of two different anomalous diffusion processes.
Nonetheless, RANDI manages to achieve performances that are comparable with those obtained on trajectories coming from a single process.
For the classification of the segments before the switch, our method achieves an accuracy of $68\%$, which is just below the performance that would be achieved by the networks used for the classification of trajectories without a switch on trajectories of  similar length. Indeed, for trajectories without switching of length from $3$ to $199$, we would achieve an accuracy of $70\%$. The confusion matrix for the results of the first segment is shown in Fig.~\ref{fig:fig5}(b). We observe that CTRW and LW are more successfully identified while ATTM, FBM and SBM display a lower accuracy. These results are also similar to the results we have seen in the classification tasks (Fig.~\ref{fig:fig2}(b)). For the classification of the segments after the switch, RANDI has an accuracy of $64\%$. We see that the second segment is more difficult to classify.
A possible explanation is that, for a non-stationary process it is important to know for certain the starting point. The fact that for the second segment there is some additional uncertainty about when the segment actually starts then makes classification more difficult.
The confusion matrix for the classification of the second model (after the switch) is shown in Fig.~\ref{fig:fig5}(c). 
The prediction of the switching time vs the ground truth is shown in Fig.~\ref{fig:fig5}(d).
We see that the method is very accurate in identifying switching points that do not occur too early or too late. Since trajectories with early or late switches look similar to trajectories with no switches at all the network sometimes mixes them up and mistakes an early  switch for a late one (and vice versa). 
This feature is not accidental and is a consequence of the fact that in the training procedure we minimize the distance between the predicted and the true switching time in a periodic fashion as discussed in the section \ref{sec:task3_train}.
For our submission to the AnDi challenge
to mitigate this issue we average the predictions of two slightly different networks that are trained independently.
Here, for the sake of clarity we do not perform this averaging.
The resulting MAE for the switching time is $14$ time steps and the corresponding root-mean-square error (RMSE) is $41$ time steps. 
The MAE is less sensitive than the RMSE to mistaking an early switch for a late one (or vice versa).
Their large difference confirms the intuitive picture that the RMSE is dominated by errors in predictions that swap early and late switching times.
For the inference of the anomalous exponent we achieve a MAE of $0.29$ 
for the first segment and $0.31$ 
for the second one. Again, this is just above the one we would obtain using the nets trained for the inference of a trajectories of length $2$ to $199$ with fixed exponent, which is $0.27$. This indicates that our method is very robust with respect to the additional uncertainty induced by the switching of trajectories since we recover performances comparable to the ones for trajectories with constant exponents and model types. The distributions of the predictions of anomalous diffusion exponents for the first and second segment are shown in Fig.~\ref{fig:fig5}(e) and Fig.~\ref{fig:fig5}(f), respectively. We remark that with these results, RANDI was the top performing method also for the segmentation task in the AnDi challenge.  

An alternative strategy, which is often pursued for algorithmic approaches, is to first identify the switching point and then apply methods  for inference or classification
of trajectories in which the model does not change.
We found that using our networks, this strategy did not reach the same performances.
This suggests that RANDI manages to limit the propagation of the uncertainty in when the change took place to the tasks of inference and classification.

\subsection{Higher dimensional data}
For higher dimensional data, we train RNNs using the trajectories of the corresponding dimension. Again, we train one RNN for the joint inference of exponents and switching time and two RNNs for classification of the first and second segment, respectively.
In two dimensions we reach an accuracy of $72\%$ for the first segment and $67\%$ for the second one. The MAE for the exponent inference is $0.25$ for the first segment and $0.28$ for the second one. The RMSE (MAE) for the switching time is $30$ time steps ($8$ time steps).
In three dimensions classification is performed with $82\%$ accuracy for the first segment and $74\%$ for the second one. The MAE for the exponent inference is $0.22$ for the first segment and $0.25$ for the second one. The RMSE (MAE) for the switching time is $24$ time steps ($5$ time steps).

\subsection{Training procedure}\label{sec:task3_train}
Since it is difficult to combine inference and classification in a single architecture, we train different networks for this task. We train 3 different networks for analyzing trajectories for segmentation, the details of the architectures for these networks are shown in Table~\ref{table1}. One network is trained to classify the anomalous diffusion model class of the first segment and a second network to classify the  second segment. One network is trained for the inference tasks to predict the switching time and the anomalous diffusion exponents ($\alpha_1$ and $\alpha_2$) of both segments. A trajectory that features an early switch looks very similar to a trajectory with a late switch (and to a trajectory with no switch at all). In order to facilitate the learning process we consider the switching time in a periodic fashion.
More precisely, we use as labels (and outputs)  for the switching time the $\sin(2\pi t_s/200)$ and $\cos(2\pi t_s/200)$.
The switching time is then obtained by inverting the relation by means of the arctan function.
The network takes as inputs the increments $\Delta x_i$  with a zero added at the end of the trajectory. These inputs are reshaped into blocks of size $2$, $4$ and $6$ for trajectories of dimension one, two and three, respectively.

\section{Conclusions}

With this study we have shown that RNNs are an excellent tool for extracting information about anomalous diffusion even from single, short and noisy trajectories.
We have shown how stacking two LSTM layers (an architecture proposed in \cite{Bo2019} for the study of fractional Brownian motion) can successfully deal with different models of anomalous diffusion, extracting several relevant parameters (diffusion models, anomalous exponents, and switching times), and can easily be made robust against noise even without knowing the underlying model of anomalous diffusion.
Of course, a priori knowledge about the type of anomalous diffusion or of the noise level greatly simplifies the task. In case this knowledge is available,
it is advisable to train the RNNs on data the resemble as closely as possible the experimental situation of interest.

A striking feature of our approach is that the same basic architecture, with minor modifications mainly in the output layers, achieves excellent results for very different tasks  that varied from inference to classification and change-point detection, and for trajectories of different lengths and spatial dimensions.
Indeed, RANDI outperforms the best-scoring methods that participated in the Anomalous Diffusion Challenge  in $8$ of the $9$ proposed tasks \cite{Munoz-Gil2020}.
Together with other studies \cite{Bo2019,Argun2020}, these findings indicate that LSTM are a truly versatile and robust tool for the analysis of single short stochastic time series 
and suggest that they might be suitable for the studies of even more complicated systems.

\begin{figure}[h t]
\setcounter{figure}{0}   
\renewcommand{\figurename}{Supplementary Figure}
	\begin{center}
		\includegraphics[width=\textwidth]{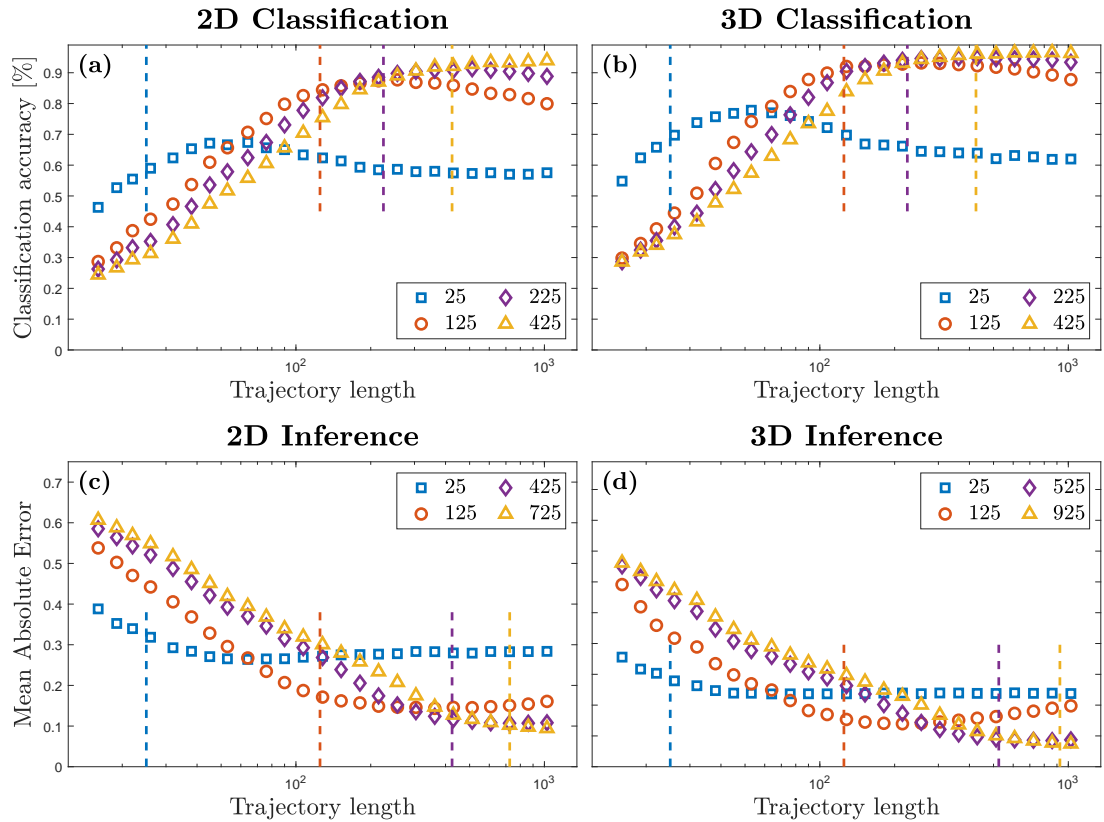}
	\end{center}
	\caption{{\bf Performance of the neural networks on trajectories with different length for higher dimensions.} 
	{\bf (a)} Classification performance of RNNs that are trained with trajectories of length 25 (blue squares), 125 (red circles), 225 (purple diamonds) and 425 (gold triangles) are shown as a function of input length of the validation data for 2D data.
	{\bf (b)} Classification performance of RNNs that are trained with trajectories of length 25 (blue squares), 125 (red circles), 225 (purple diamonds) and 425 (gold triangles) are shown as a function of input length of the validation data for 3D data. 
	{\bf (c)} Inference performance of RNNs that are trained with trajectories of length 25 (blue squares), 125 (red circles), 425 (purple diamonds) and 725 (gold triangles) are shown as a function of input length of the validation data for 2D data.
	{\bf (d)} Inference performance of RNNs that are trained with trajectories of length 25 (blue squares), 125 (red circles), 525 (purple diamonds) and 925 (gold triangles) are shown as a function of input length of the validation data for 3D data.
	The length of trajectories used as training data for each network is shown by dashed lines in the corresponding color. }
	\label{fig:suppfig}
\end{figure}

\subsection*{Acknowledgements}
The authors wish to thank Erez Aghion and Philipp Meyer for insightful discussions, and Erez Aghion for valuable comments on the manuscript. A.A. and G.V. acknowledge support from the European Research Council (ERC) Starting Grant ComplexSwimmers (677511).

\bibliography{andi}

\end{document}